\documentclass[preprint,aps,pra]{revtex4-1}
\usepackage{amsmath,amssymb}
\usepackage{graphicx}%Include figure files
\usepackage{dcolumn} % Align table columns on decimal point
\usepackage{bm}% bold math
\usepackage{mathrsfs}
\usepackage[colorlinks,allcolors=blue]{hyperref}
\graphicspath{{figures/}}
\usepackage[colorlinks]{hyperref}
\usepackage{appendix}

\begin{document}

\title{Fast generation of Cat states in   Kerr nonlinear resonators via optimal adiabatic control }

\author{Jiao-Jiao Xue$^{1}$, Ke-Hui Yu$^{1}$, Wen-Xiao Liu$^{2}$, Xin Wang$^{1}$  and Hong-Rong Li$^{1}$}

\email{hrli@xjtu.edu.cn}

\affiliation{$^{1}$Institute of Theoretical Physics, School of Physics, Xi'an Jiaotong University, Xi$^{\prime}$an 710049, China\\$^2$Department of Physics and Electronics, North China University
of Water Resources and Electric Power, Zhengzhou 450046,
People’s Republic of China}

\begin{abstract}
Macroscopic  cat states  have been  widely studied to illustrate fundamental principles  of quantum physics as well as    their applications  in  quantum information processing. In this paper, we propose a quantum speedup method for adiabatic creation of cat states in  a  Kerr nonlinear resonator  via  optimal adiabatic control. By  simultaneously  adiabatic tuning  the cavity detuning and driving field strength, the width of  minimum energy gap between the target trajectory and non-adiabatic trajectory can be widen, which allows us to speed up  the evolution along the adiabatic path. Compared with the previous proposal of preparing the cat state  by  only controlling two-photon pumping strength in a Kerr nonlinear resonator, our method can prepare the target state with much shorter time, as well as  a high fidelity and a large non-classical volume. It is worth noting that the cat state prepared by our method is also robust against single-photon loss very well. Moreover, when our proposal has a large initial detuning, it will  creates a large-size cat state successfully. This proposal of preparing cat states can be implemented in superconducting quantum circuits, which provides a quantum state resource for quantum information encoding and fault-tolerant quantum computing.
\end{abstract}

\maketitle

\section{Introduction}
Schr{\"o}dinger's cat states, i.e., quantum superpositions of two macroscopically distinct states, play an important role in understanding the  fundamental quantum theory \citep{zurek2003decoherence,haroche2013nobel,leggett1985quantum,yurke1986generating,roy1991tests,kim1992schrodinger,
shimizu2002stability,shimizu2005detection,morimae2006visualization,morimae2010superposition,frowis2012measures,
arndt2014testing,frowis2015linking,jeong2015characterizations,fischer2015photonic,abad2016scaling,lambert2016leggett,haroche2013nobel}. More recently, it has been recognized that cat states are also a useful resource for quantum information processing \citep{puri2019stabilized,mirrahimi2014dynamically,
wang2016schrodinger,michael2016new,goto2016universal,wang2007nonadiabatic,ma2021quantum,
puri2020bias,xu2021engineering,puri2020bias}. As an alternative for using two-level systems to realize quantum information encoding,  cat qubits are embedded in the infinite-dimensional Hilbert space, which can  encode   quantum information redundantly without  introducing  additional decay channels \citep{mirrahimi2014dynamically,ma2021quantum}.
In additional, stabilized cat codes can significantly reduce the resource overhead for fault-tolerant quantum computing, through a set of
 bias-preserving gates  with a biased noise channel \citep{xu2021engineering,puri2020bias}. Therefore, the ability to reliably create and manipulate cat states is  highly desirable.

Up to now, cat states have been generated by various approaches \citep{vlastakis2013deterministically,leghtas2013deterministic,wang2016schrodinger,leghtas2013hardware,leghtas2015confining,ourjoumtsev2007generation,zhan2020preparing,takahashi2008generation,ourjoumtsev2006generating, wineland2013nobel,sackett2000experimental,leibfried2005creation,hatomura2018shortcuts,friedman2000quantum,van2000quantum,chen2021shortcuts}. Some of these  methods are based on  unitary evolution, which relies on  strong  dispersive interaction induced by  the ancillary system and field to transfer an arbitrary state of the ancillary system into a cat state \citep{leghtas2013deterministic,vlastakis2013deterministically,wang2016schrodinger}.  Other methods are based on  quantum reservoir engineering, which employs  a two-photon  dissipation process to ensure that the steady state of the system is  cat state \citep{leghtas2013hardware,leghtas2015confining,wang2014preparing,wang2017hybrid,li2012dissipative,wang2013reservoir,qin2021generating,zhou2021generating}. 
However, these two types of methods are respectively sensitive to environmental noise and single-photon loss, high-fidelity preparation and manipulation of cat states is still challenging.

Kerr-nonlinear resonators (KNR) have attracted much interest   due to its  rich physical  characteristics \citep{miranowicz1990generation,gu2017microwave,dykman2012fluctuating,siddiqi2005direct,castellanos2008amplification,munro2005weak,moloney2018nonlinear,yin2012dynamic,zhao2018two,goto2016bifurcation,goto2019demand,grimm2020stabilization,grimm2020stabilization,teh2020dynamics,
sun2019schrodinger,lu2013quantum}. The single-photon Kerr regime has been realized in superconducting quantum circuits \citep{kirchmair2013observation},  with
$K/\kappa \sim 30$ demonstrated experimentally (where $K$ is Kerr nonlinearity strengh and $\kappa $ is single-photon loss). Rencently, a method of preparing  cat states in a KNR by the use of a two-photon driving has been proposed \citep{puri2017engineering}. This method takes advantages of the fact that even and odd cat states are  degenerate ground state of KNR  under a two-photon driving. Even in the presence of  single-photon loss, the  high-fidelity cat states can be  produced by this method. Since the preparation of cat states using this method relies on adiabatic two-photon driving,  a long evolutionary time is required for this method. How to shorten the preparation time of  cat states in  KNR without affecting the fidelity  is worth exploring.

To overcome long-time cost problem of adiabatic evolution, we propose a quantum speedup method by tuning one more additional detuning parameter $\Delta$. In the parametric space $\{\Delta,\beta\}$, an optimal evolution path can be
numerically found by applying gradient-descent method. Along the optimal path, the KNR can be steered into the target state. Compared to previous method with only one parameter $\{\beta(t)\}$, the state preparation time of our  method is much shorter as well as a high fidelity. Due to the existence of time-dependent extra parameter $\{\Delta(t)\}$, we dynamically enhance the state preparation mechanism. We then analytically illustrate  this speed up mechanism:  enhancing the minimum energy gap $\Delta_{\text{min}}$  between the target trajectory  and non-adiabatic trajectory. The minimum energy gap $\Delta_{\text{min}}$ is inversely proportional to the adiabatic evolution time $T$, which  increasing $\Delta_{\text{min}}$ can shorten  $T$.  Significantly, our proposal also retains the advantages of previous only controlling one parameter $\{\beta\}$ method in Ref. \citep{puri2017engineering}, which  resists single-photon loss very well.  Moreover, when the initial detuning $\Delta_0$ of  our method takes a larger value,  the large-size cat states can be created. 

This paper is constructed as follows. In section \ref{hh}, we introduce the  eigen spectrum of KNR  with and without two-photon drive. The basic properties of  KNR, which are necessary for our discussions, are reviewed.  We summarize the optimal adiabatic control method for preparing cat states   in section \ref{ai}.  In section \ref{ll}, we demonstrate generation of the cat state via  optimal adiabatic control sequences.  We first take the initial detuning $\Delta_0 = 2K$ and initial drive strength $\beta_0 = 0$ as   an example to show the optimal path to prepare   cat states, and  then compare it with the method of preparing the cat state  with only one parameter $\{\beta\}$.  In section \ref{yff}, we discuss the effect of different  initial detuning $\Delta_0 $ on our preparation of cat states proposal.  In section \ref{doujiang}, we discuss a possible implementation of our protocol using a superconducting quantum circuit.  We summarize in section  \ref{yangsheng}.

\section{eigen spectrum of Kerr  nonlinearity resonator}\label{hh}
We consider a KNR with resonant frequency $\omega_c$ driven by a field with   frequency $\omega_p$ \citep{marthaler2007quantum}. In the rotating frame of the driving field, the  Hamiltonian of the system is
\begin{align}\label{eq1}
H = K a^{\dag}a^{\dag}aa +\Delta a^{\dag}a - \beta(a^{\dag 2} + a^2),
\end{align}
where $a$ denotes the annihilation operator of photons in the cavity, $K$ is the Kerr coefficient, which restricts  $K > 0$ in this work, and $\Delta = \omega_c-\omega_p/2 $ is the cavity detuning. Here, $\beta$ denotes the strength of the cavity-driving field. Without loss of generality, we assume $\beta$ to be a non-negative real number.
\begin{figure}[!htbp]
	\centering
	\includegraphics[scale=0.38]{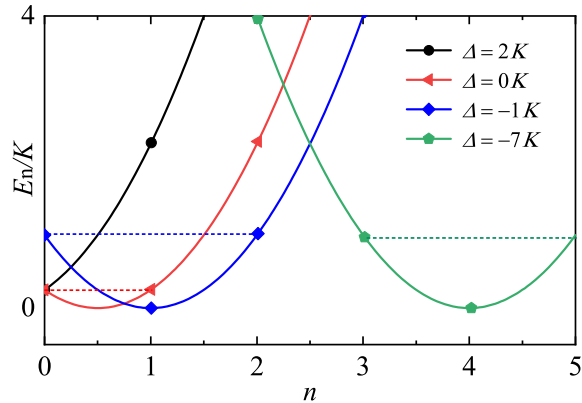}
\caption{Eigenenergies of the Hamiltonian in Eq. (\ref{eq1}) plotted as a function of photon number $n$ for drive $\beta = 0$. The curves from left to right correspond to $\Delta/\text{K} = 2$ (black), $0$ (red), $-1$ (blue), and $-7$ (green). The values of the energies are represented by symbols, which refer to integer value of $n$. The dashed lines are intended to show the degeneracy : $E_0 = E_1$; $E_0 = E_2$; $E_3 = E_5$.}	
    \label{fig1}
\end{figure}

In the absence of the drive ($\beta =0$), the Hamiltonian in Eq. (\ref{eq1}) can be diagonalized  in the basis of Fock states \citep{zhang2017preparing}, and the eigenvalues $E_n$ can be written in the following
\begin{align}\label{eq2}
E_n = \bar{E}_n - \bar{E}_0,\ \bar{E}_n = K\left(n + \frac{\Delta}{2K} -\frac{1}{2}\right)^2.
\end{align}
From Eq. (\ref{eq2}), for the integer $n$, $E_n$ is a simple parabola with a minimum at $n = 1/2 -\Delta/2K$ (see Fig.(\ref{fig1})). For $\Delta > 0$, there is no degenerate in KNR.  For $\Delta \leqslant 0$, the symmetrical eigenstates of $n = 1/2 -\Delta/2K$ will degenerate. Therefore, the ground state and degenerate states of KNR can be changed by adjusting the detuning $\Delta$.
\begin{figure}[!htbp]
	\centering
	\includegraphics[scale=0.17]{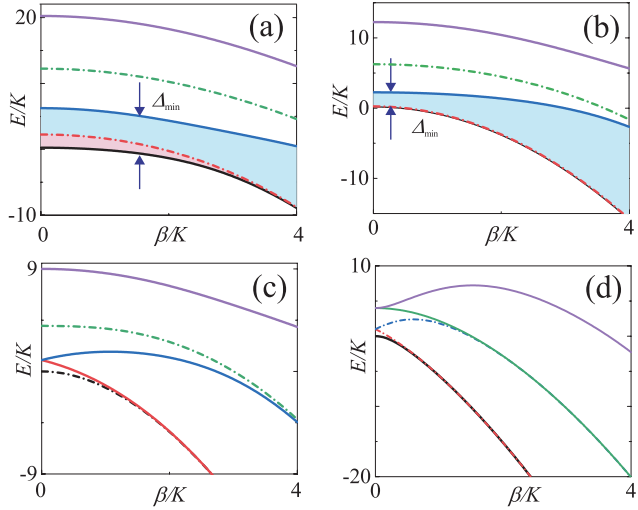}
    \caption{Eigenenergies of the Kerr nonlinearity resonator (KNR)  plotted as functions of driving $\beta$ for different detuning $\Delta$. (a) $\Delta = 0$, (b) $\Delta = 2K$, (c) $\Delta = -K$, (d) $\Delta = -7K$.  The Hamiltonian energy levels of even and odd parity are represented by solid and dashed lines, respectively.  }
\end{figure}

For different detuning $\Delta$, the  eigen energies of the KNR  versus   $\beta$ are shown  in Fig. 2. A common trend is that,  with large $\beta$, the Hamiltonian energy levels of the same parity repel each other, whereas  adjacent energy levels with opposite parity attract each other and eventually merge together. When $\Delta \geqslant 0$, the vacuum state $|0\rangle$ and the first Fock state $|1\rangle$ are the two lowest energy states of the system, see Fig. 2(a) and  Fig. 2(b). As the driving $\beta$ increases, they will degenerate to a new ground state.  During the driving process, the  minimum level gap $\Delta_{\text{min}}$ between the ground state and the first excited state  is determined by  $\Delta$. With  increasing the detuning $\Delta$, the  minimum level gap $\Delta_{\text{min}}$ will also increases.

When $\beta$ becomes larger than $\Delta$, the  eigenstates  of the system  form many two-dimensional nearly degenerate subspace spanned by $\{D(\pm \alpha)|n\rangle\}$ for each $n$, where $ \pm\alpha = \pm \sqrt{\beta/K}$  are the locations of the local maxima \citep{zhang2017preparing}. The  eigenstates of the system are  two quasi-orthogonal states   $|\psi^{\pm}_{n}\rangle = N^{\pm}[D(\alpha)\pm D(-\alpha)]|n\rangle$ with $N^{\pm} = 1/\sqrt{2(1\pm e^{-2\alpha^2})}$,  which are the even- and odd-parity states \citep{puri2019stabilized,chamberland2020building},  marked  by the red and blue levels respectively in  Fig. 3. The cat states $|C^{\pm}_{\alpha}\rangle = N^{\pm}(|\alpha\rangle \pm |-\alpha\rangle)$ are degenerate ground state  of the KNR in the limit of large $\beta$.  As shown in Fig. 3, this  degenerate cat state subspace (green) is separated from the rest  of Hilbert space (orange) by a large energy gap $\Delta_{\text{gap}}$, which is well approximated as $\Delta_{\text{gap}} \sim 4K|\alpha|^2$ \citep{puri2020bias}. For $\beta = 0$, $|C^{+}_{\alpha}\rangle$ and $|C^{-}_{\alpha}\rangle$ are the vacuum state $|n = 0\rangle$ and the first Fock state $|n = 1\rangle$, respectively. Therefore, we can achieve the transition from the Fock state to the cat state by increasing the drive strength $\beta$.

\begin{figure}[!htbp]
	\centering
	\includegraphics[scale=0.22]{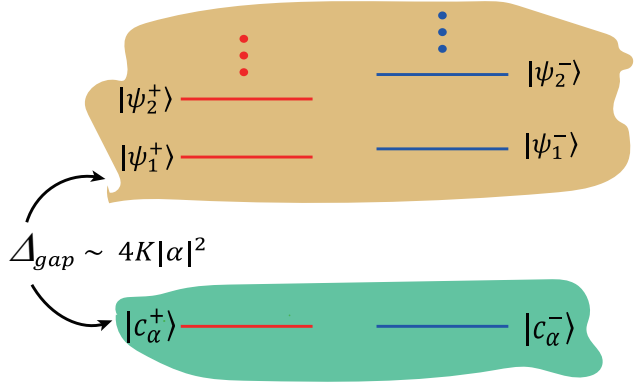}

    \caption{Eigenspectrum of the  KNR in the limit of $\beta \gg \Delta,$ $K$. The eigenspectrum can be divided into an even (red) and an odd parity (green) manifold. The cat subspace (green) is separated from the rest of  excited states (orange) by an energy gap $\Delta_{\text{gap}} \sim 4K|\alpha|^2$. }

\end{figure}

\section{Optimal Adiabatic control  for preparing cat states}\label{ai}
In this section, we introduce the   optimal adiabatic control method to prepare   cat states in the KNR. This method is based on quantum adiabatic theory, we first give a brief description of  the QAT  \citep{shankar2012principles,aharonov1987phase,wu2005validity,guery2019shortcuts}. If a system Hamiltonian is given by $H(C(t))$, where $C$ is some external coordinate which changes slowly and appears parametrically in $H$. The QAT shows that the probability of the system transition from the initial state $|m(0)\rangle$ to all $|n(t)\rangle (m \ne n)$ can be ignored, and the system will remain in the instantaneous eigenstate $|m(t)\rangle$ corresponding to the initial state.  In this process, the  quantum adiabatic condition
\begin{align}\label{xxx}
\left | \frac{\langle m(t)|\frac{\partial}{\partial t}| n(t)\rangle}{E_m(t) - E_n(t)} \right | \ll  1
\end{align}
must be satisfied, which means that the non-adiabatic transition between energy levels must be slow compared to the inverse energy gap between them. In Ref. \citep{aharonov1987phase}, the  condition of the QAT is also summarized as
\begin{align}\label{mgg}
\left |\frac{\langle m(t)|\dot{H}| n(t)\rangle}{(E_n(t) - E_m(t))^2}\right | \ll  1,
\end{align}
which is  equivalent to Eq. (\ref{xxx}). The dot in Eq. (\ref{mgg}) denotes the differentiation with respect to parameter $C$ and time $t$. The system adiabatic evolution time $T$ can be estimated from \citep{aharonov1987phase}
\begin{align}
T \gg \frac{1}{\Delta_{\text{min}}},
\end{align}
where $\Delta_{\text{min}}$ is the  minimum energy level difference between the energy levels of the non-adiabatic transition. Therefore, in order to shorten the adiabatic evolution time $T$, it is necessary to increase the minimum energy level gap $\Delta_{\text{min}}$ of the system.

Then we give the initial conditions of the  optimal adiabatic control method based on QAT.  The system is initialized in the vacuum state $|0\rangle$ with initial drive strength $\beta_0 =0$, and   vary $\Delta$ and $\beta$  simultaneously to prepare the even cat state. The preparation of the odd cat state is same as the even cat state, except that the initial state of the system is prepared in Fock state $|1\rangle$. In the process of preparing  the cat state, we should adiabatically keep the system in the ground state at all times to satisfy the adiabatic condition.  According to  the energy level structure of the KNR in section \ref{hh}, in the process of driving, the  minimum energy gap $\Delta_{\text{min}}$  between the ground state and higher excited states decreases with  increasing the initial detuning $\Delta$, as shown in  Fig. 2(a) and 2(b).  Therefore, we set the initial detuning $\Delta > 0$ and assume it is time-dependent. In our proposal, $\beta$ ($\Delta$)  increases (decreases) with time during the preparation of cat states.

Next, we  determine the exact shape of the optimal adiabatic path. We parametrize the path as $C = \{\Delta(s), \beta(s) : s  \in [0,S]\}$, where $S$ is the total length of the path and define the small step length $ds$ in $(\Delta, \beta)$ space  as $ds = \sqrt{(d\Delta)^2 + (d\beta)^2}$.  The trajectories of the parameters are obtained by introducing time dependence $s(t)$, which represents the distance covered up to time $t$.  According to the condition for adiabaticity, an instantaneous penalty function $P_C(t)$ characterizing the probability of leaving the ground state is given by \citep{yanagimoto2019adiabatic}
\begin{align}\label{chikui}
P_C(t) = \sum_{n \ne 0} \frac{|\langle \phi_n(t)|\frac{dH}{dt}|\phi_0(t)\rangle|}{|E_n(t) - E_0(t)|^2} = \frac{ds}{dt}Q_C(s).
\end{align}
Note that the time-dependent  parameters in Eq. (\ref{chikui}) are given via $s(t)$. Here, the function $Q_C$ is  \citep{yanagimoto2019adiabatic}
\begin{align}\label{elm}
Q_C(s) = \sum_{n \ne 0}\frac{|\frac{d\beta}{ds}L_n(s) + \frac{d\Delta}{ds}M_n(s)|}{|E_n(s) - E_0(s)|^2},
\end{align}
with $|\phi_n(s)\rangle = |\phi_n(\Delta(s), \beta(s))\rangle$ and
\begin{align}
&L_n(s) = \langle \phi_n(s)|(a^{\dag 2} +a^2)| \phi_0(s)\rangle,\\&
M_n(s) = \langle \phi_n(s)|a^{\dag}a| \phi_0(s)\rangle.
\end{align}
The total penalty of the path $C$ can be calculated by
\begin{align}\label{xf}
I[C] = \int_0^TdtP_C(t) = \int_0^SdsQ_C(s),
\end{align}
where $T$ is the total time to traverse the  path,  so that $s(T) = S$.  We then  use the gradient-descent method to find  a set of points that minimizes the second integral in the limited parameter space $\{\Delta,\beta\}$. By minimizing the second integral, we can determine the optimal path without preknowledge of $s(t)$.
 
Finally we choose the time dependence $s(t)$ to determine the optimal control sequence after obtaining an optimal path $C$. The penalty function $P$ is a non-negative function of time, so by Eq. (\ref{xf}),
\begin{align}\label{moyu}
\max_{t} P_C(t) \geqslant \frac{I[C]}{T}.
\end{align}
Choose a parametrization $P_C(t) = I[C]/T$,   the inequality  in Eq. (\ref{moyu}) is always saturated \citep{yanagimoto2019adiabatic}. Under this choice, we have
\begin{align}\label{kk}
t(s) = \frac{T}{I[C]}\int_0^s ds^{\prime}Q_C(s^{\prime}).
\end{align}
Through the  numerically inverting $t(s)$, we  obtain $s(t)$, which in turn generates the time-dependent trajectories of parameters $\beta(t)$ and $\Delta(t)$ and gives the total desired time $T$. Therefore, the optimal adiabatic control sequences  for preparing cat state are obtained.

\section{CAT-STATE GENERATION}\label{ll}
\begin{figure}[!htbp]
	\centering
	\includegraphics[scale=0.36]{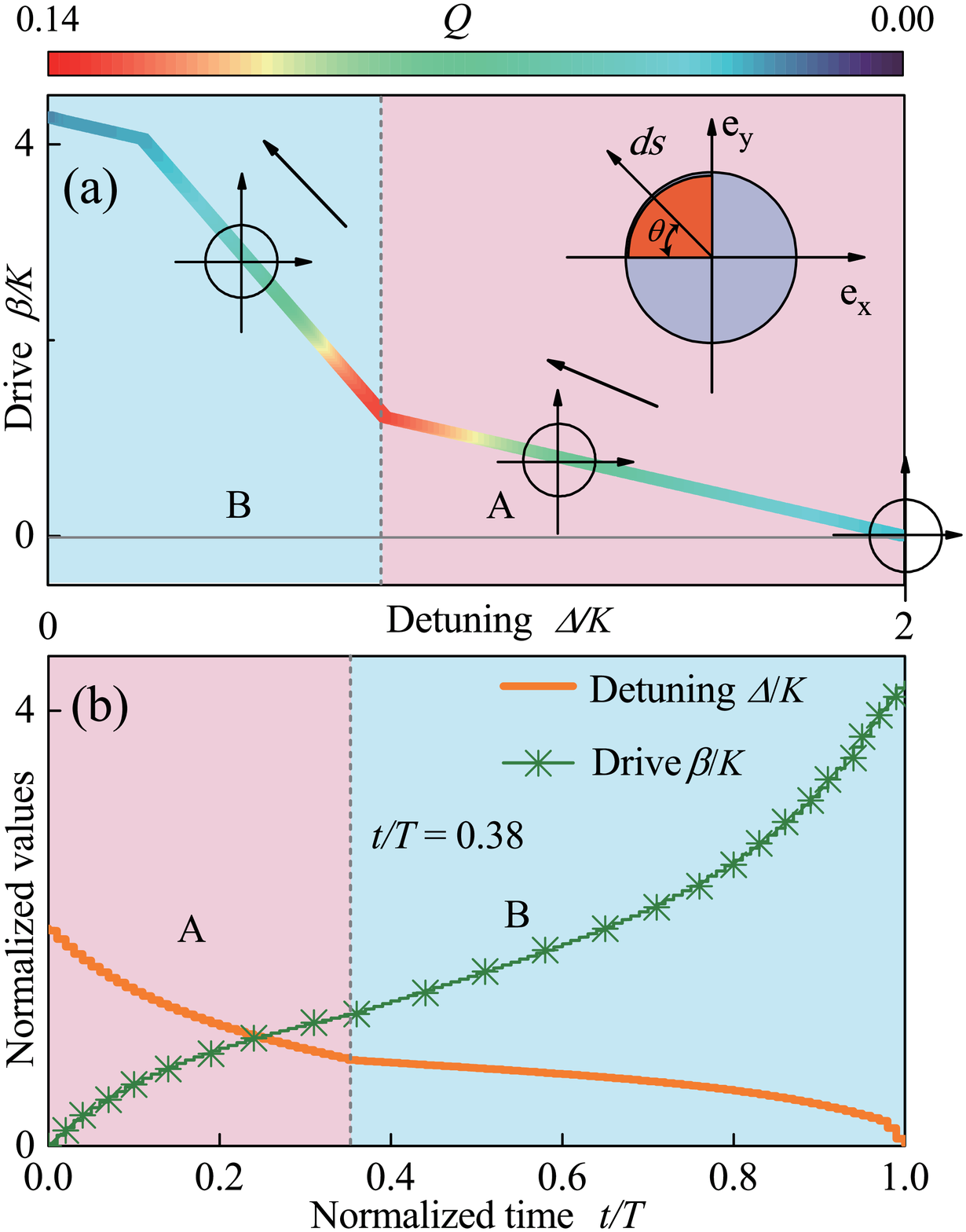}
    \caption{(a)By setting initial detuning $\Delta_0 = 2K$ and  initial drive strength $\beta_0 = 0$ to generate the optimal path for preparing  cat states. The color of path indicates the penalty $Q$ at each point. Arrows indicate the direction of gradient-descent. (b) Drive $\beta(t)$ and detuning $\Delta(t)$ as prescribed by Eq. (\ref{kk}) versus as $t$. Different background colors are used to represent the A and B regions. }	
    \label{fig3}
\end{figure}

In this section,  we demonstrate generation of  cat states via  optimal adiabatic control method and show the advantages of this method. We initialize the system in the vacuum state $|0\rangle$ or the first Fock state $|1\rangle$ with initial detuning $\Delta_0 = 2K$ and initial drive strength $\beta_0  = 0$. For convenience,  we set $K=1$ to determine the timescale in the following discussion. As shown in Fig. 4(a), we  introduce the parameter $\theta$ and vector plane $(\vec{\mathbf{e}}_x,\vec{\mathbf{e}}_y)$,   the  small step with direction $d \vec{s}$ can be described by
\begin{align}
\begin{aligned}
d \vec{s} &= ds \ \text{cos} \theta \  \vec{\mathbf{e}}_x + ds \ \text{sin} \theta \ \vec{\mathbf{e}}_y \\&= d\Delta \ \vec{\mathbf{e}}_x + d\beta \ \vec{\mathbf{e}}_y,
\end{aligned}
\end{align}
where $\theta \in (\pi/2, \pi)$ is determined by the changing trend of $\Delta$ and $\beta$. The parameters  $\Delta$ and $ \beta$ can be  expressed as:
\begin{align}
\begin{aligned}
\Delta &= \Delta_0 + d\Delta  = 2 + ds \ \text{cos} \theta\\
\beta &= \beta_0 + d\beta  = 0 + ds \ \text{sin} \theta.
\end{aligned}
\end{align}
To find the optimal route in parametric space $\{\Delta, \beta\}$, we need to determine $\theta$ corresponding to each step $ds$. This can be achieved by a numerical calculation  based on gradient-descent method. Specifically, we first choose $(2,0)$ as the starting point, and make a circle with $(2,0)$ as the center and $ds$ as the radius. Then, we substitute all the points on the second quadrant of the circle into Eq. (\ref{elm}), and  get a set of $Q_C(s_0 + ds)$. We choose a $\theta$,  which satisfies
\begin{align}
\text{min}[Q_C(s_0 + ds)- Q_C(s_0)],
\end{align}
as the direction of gradient-descent.  The position $s_0 + ds$  is the point in the optimal path. We repeat the above process with the found point as the new starting point, until $\Delta$ is reduced to $0$ to stop the above process. The optimized path for preparing cat state with initial detuning $\Delta_0 = 2K$ and $\beta = 0$ as shown in Fig. 4(a), the penalty function $Q_C$ at each point in our path satisfies the condition of adiabatic approximation $Q_C \ll 1$. We divide the optimized path into two regions according to the changing trend of penalty function $Q_C$ at each point, named A and B. The penalty function $Q_C$ for each point gradually increases in area A, and gradually decreases in area B.

To better understand  the dynamics of this process, we use a displacement operator $D(\alpha)$ in our Hamiltonian of Eq. (\ref{eq1}), where $D(\alpha) = \text{exp}(\alpha a^{\dag} - \alpha^{\ast} a)$. Under a displacement transformation $D(\alpha)$, the system Hamiltonian becomes \citep{puri2017engineering}
\begin{align}
\begin{aligned}
H^{\prime} = & D^{\dag}(\alpha)HD(\alpha)\\ = &[(2K \alpha^2 \alpha^{\ast} - 2\beta \alpha^{\ast} + \Delta \alpha)a^{\dag} + \text{H.c}]  \\& +[(K \alpha^2 - \beta)a^{\dag 2}+ \text{H.c}] + (4K|\alpha|^2 + \Delta)a^{\dag}a \\&+ K a^{\dag 2}a^2 + (2K\alpha a^{\dag 2}a + \text{H.c}),
\end{aligned}
\end{align}
where we have dropped the constant term $E = 4K|\alpha|^4 - \beta(\alpha^2 + (\alpha^{\ast})^2)+\Delta |\alpha|^2$ that represents a shift in energy. We take $\alpha$ to satisfy \citep{puri2017engineering}
\begin{align}\label{tashanhe}
2K \alpha^2 \alpha^{\ast} - 2\beta \alpha^{\ast} + \Delta \alpha = 0,
\end{align}
such as to inhibit the the unwanted single-photon pumping. The Hamiltonian $H^{\prime}$ now  reads
\begin{align}\label{dafengchui}
\begin{aligned}
H^{\prime} = &[(K \alpha^2 - \beta)a^{\dag 2}+ \text{H.c}] + (4K|\alpha|^2 + \Delta)a^{\dag}a \\&+ K a^{\dag 2}a^2 +  (2K \alpha a^{\dag 2}a + \text{H.c}).
\end{aligned}
\end{align}
Equation (\ref{tashanhe}) gives $\alpha = 0, \pm \sqrt{(2\beta-\Delta)/2K}$, where we assume $\alpha = \alpha^{\ast}$.

For $\alpha = 0$,  the  first two terms of  Eq. (\ref{dafengchui}) represent a nearly resonant parametric drive of  strengh $\beta$ \citep{puri2017engineering}. When $\beta \ll \Delta$, the eigenstate of the system is approximately regarded as  a Fock state.  As the detuning $\Delta$ (drive strength $\beta$) continues to decrease (increase), the difference between $\Delta$  and $\beta$  decreases. However, the displaced amplitude $\alpha$ still very small, the instantaneous eigenstate of the system is still approximately regarded as a Fock state. The energy gap between the ground state  and higher excited states is dominated by  detuning $\Delta$. As  $\Delta$ decreases, the probability of the ground state transitioning to  higher excited states gradually increases. Therefore, the penalty function $Q_C$ that  measures the transition probability will gradually increases with $\Delta$  decreases. The changing trend of the penalty function $Q_C$ is consistent with that in region A, which will increases with  $\Delta$.

For $\alpha = \pm \sqrt{(2\beta-\Delta)/2K} $, the displaced Hamiltonian can be rewritten as \citep{puri2017engineering}
\begin{align}\label{hbhll}
\begin{aligned}
H^{\prime} &= [(-\frac{\Delta}{2})a^{\dag 2}+ \text{H.c}] + (4K|\alpha|^2 + \Delta)a^{\dag}a \\&+ K a^{\dag 2}a^2 + (2K\alpha a^{\dag 2}a + \text{H.c}).
\end{aligned}
\end{align}
The first two terms of Eq. (\ref{hbhll}) now represent a parametric drive  whose amplitude has an absolute value of $\Delta/2$, and is detuned by $4K|\alpha|^2 + \Delta$. When $2\beta \gg \Delta$, the  eigenstates of the system are spanned by $\{D(\pm \alpha) |n\rangle\}$ for each $n$, where $\pm \alpha = \pm \sqrt{(2\beta-\Delta)/2K}$. The probability of the ground state $D(+\alpha)|0\rangle$  transitioning to  other excited sates  $D(-\alpha)|n\rangle$ can be calculated by \citep{wang2019quantum}
\begin{align}\label{caomei}
\langle n |D^{\dag}(-\alpha) H D(+\alpha)|0\rangle \approx e^{-2\alpha^2} \approx e^{|(2\beta-\Delta)/K|}.
\end{align}
From Eq. (\ref{caomei}), we find that the transition probability between the ground state and other excited states will decreases with  $2\beta-\Delta$. The changing trend of the penalty function $Q_C$  in region B is consistent with Eq. (\ref{caomei}), which will decreases with  $2\beta-\Delta$.

\begin{figure}[!htbp]
	\centering
	\includegraphics[scale=0.34]{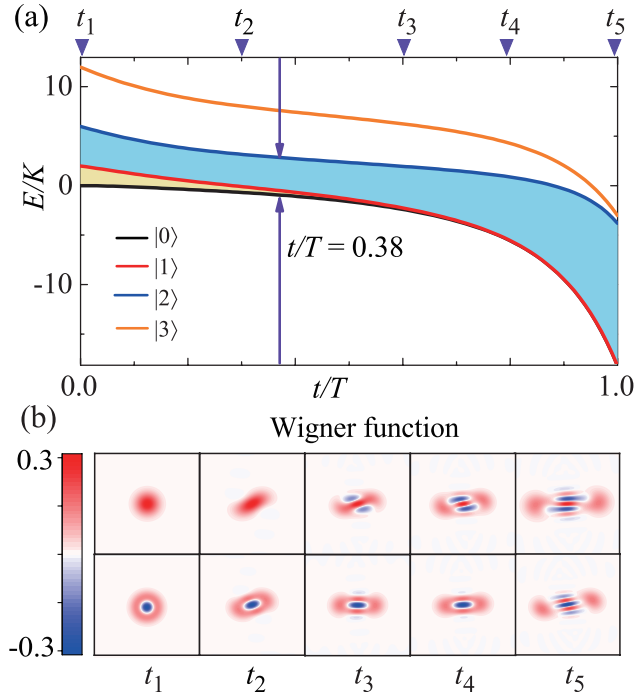}
    \caption{(a)Time evolution of the eigenenergies in the preparation of cat states adiabatically by adopting the optimal sequences $\{\Delta(t),\beta(t)\}$. (b) Wigner function $W(\alpha)$ at  times $t_1$, $\dots$, $t_5$ shown on the top of panel (a). The first and second rows correspond to the initial state is vacuum state $|0\rangle$ and Fock state $|1\rangle$, respectively. }	
\end{figure}

The optimal control sequence of detuning $\Delta(t)$ and driving strength $\beta(t)$ corresponding to each moment in the entire adiabatic evolution cycle is shown in Fig. 4(b).  The detuning $\Delta(t)$ and drive strength  $\beta(t)$ change faster in the A region, and gradually slow down in the B region. When detuning $\Delta$ decreases to zero,  numerical solution indicates that the final pulse is $\beta_f = 4.3K$. At $t = T$, the system  evolves to target Hamiltonian
\begin{align}
H = K a^{\dag}a^{\dag}aa - \beta_f(a^{\dag 2} + a^2),
\end{align}
and can be re-written as \citep{wielinga1993quantum}
\begin{align}\label{doudou}
H = K (a^{\dag 2} - \beta_f/K)(a^{2} - \beta_f/K) - \beta_f^2/K.
\end{align}
This form of the Hamiltonian illustrate that two coherent states $|\pm \alpha\rangle$ with $\alpha = (\beta_f/K)^{1/2}$, which are degenerate eigenstates of Eq. (\ref{doudou}) with energy $\beta_f^2/K$.  Equivalently, the even-odd parity states $|C_{\alpha}^{\pm}\rangle$ are also the eigenstates of  $H$.  Therefore, after an adiabatic evolution cycle $T$, we  obtain a cat state with amplitude $\sqrt{4.3} \approx 2.1$.

After acquiring the optimal control sequences  $\{\Delta(t),\beta(t)\}$, we  obtain the information of the entire evolution process. The  eigenenergies of the ground, first, second  and third instantaneous eigenstates are plotted in Fig. 5(a). Although the vacuum state $|0\rangle$ and the  first Fock state $|1\rangle$ do not degenerate at the beginning, they will merge together and become the new ground state as the evolution time increases.  However, the  level gap between the  vacuum state $|0\rangle$ or the  Fock state $|1\rangle$ and higher excited state remains open  throughout the entire adiabatic evolution cycle. This indicates that the adiabatic condition is always satisfied in the process of preparing  cat states, and the system has always been in the ground state without transition to higher excited state. The changing trend of this energy gap is consistent with the penalty function  $Q_C$ in Fig. 4(a), which  decreases first and  then gradually increases. When $t/\text{T} = 0.38$, the energy gap will reach the minimum. We define this moment as $t_{\text{min}}$, which corresponds to the junction point of region A and region B in Fig. 4(a). We plot the simulated Wigner functions  $W(\alpha)$ corresponding to the the instantaneous ground at different times  in Fig. 5(b). After $ t_{\text{min}}$, the system gradually transitions from the Fock state  to the cat state.  If the initial ground state of the system is a vacuum state $|0\rangle$, it will evolve  into an even cat state $|C_{\alpha}^{+}\rangle$ at $t = T$. If the initial ground state of the system is the Fock sate $|1\rangle$, it  will evolves to an odd cat state $|C_{\alpha}^{-}\rangle$ at $t = T$.
\begin{figure}[!htbp]
	\centering
	\includegraphics[scale=0.32]{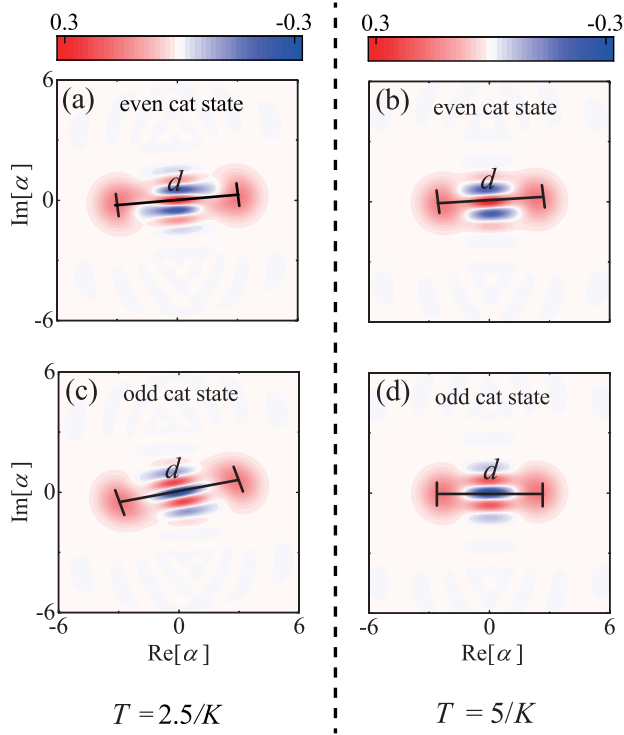}
    \caption{Wigner functions $W(\alpha)$ of  final state for a KNR  prepared by two-parameter-controlled (TPC) method and  single-parameter-controlled (SPC) method, respectively. Where (a)(c) are TPC method, and (b)(d) are SPC method. }	
\end{figure}

By tuning the two parameters of $\Delta$ and $\beta$, we successfully realize the preparation of cat states in KNR.
In order to confirm the advantages of our two-parameter-controlled (TPC) method, we compare it with the single-parameter-controlled (SPC) method.  The preparation of cat states is successfully achieved by adiabatic control of the two-photon pulsing $\{\beta\}$ in KNR in Ref.  \citep{puri2017engineering}. Different from ours, their method sets the  initial detuning $\Delta_0 = 0$, and only changes the two-photon  drive strength $\beta$ adiabatically to achieve the preparation of  cat states. The expression  of  their Hamiltonian is $H = K a^{\dag 2}a^{2} - (\beta(t)a^{\dag 2}+\beta(t)^* a^{2})$, where drive strength  $\beta(t) = \beta(t)^* = \beta_{0}[1-\text{exp}(-t^4/{T}^4)]$. When $t \gg T$, $\beta(t) \sim \beta_{0}$,  preparing the cat state  with $\alpha = \sqrt{\beta_{0}/K}$. To prepare the same target state for these two methods, we set the driving constant $\beta_0 = 4.3K$. The adiabatic evolution time for preparing cat state in  their method is $T = 5/K$, and the adiabatic evolution time of our TPC method can be calculated according to Eq. (\ref{moyu}), which is approximately  as $T = 2.5/K$.

Figure 6 shows the wigner functions $W(\alpha)$  in the   finial state of the system after their  respective adiabatic evolution cycles $T$.  Both methods achieve the preparation of  cat states, but the distance $d$ between the two coherent states $|\pm \alpha\rangle$ of the cat state prepared by our method is  much farther.  This means that  cat states prepared by our method have larger size. Moreover, we also find that the interference fringes of the cat state prepared by our method  are more clear than their method. This shows that  cat states prepared by our method are more non-classical. By comparison, the TPC method  prepares the cat state with larger size in a shorter time, and its non-classical features are better.

\begin{figure}[!htbp]
	\centering
	\includegraphics[scale=0.36]{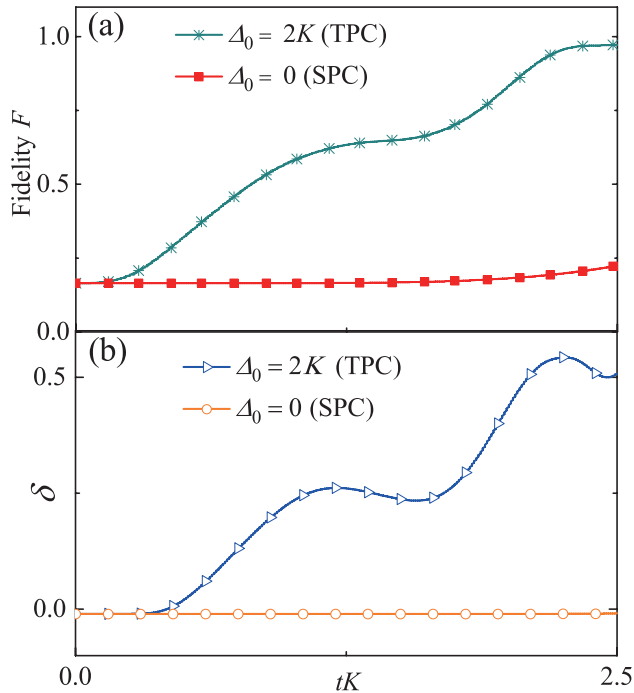}
    \caption{(Color online) Time evolution of the even cat state fidelity $F$ (a) and the nonclassicality volume $\delta$ (b) prepared by  TPC and SPC methods .}	
\end{figure}
For more convincing, we  take the even cat state as an example to compare the fidelity $F$ and the non-classical volume $\delta$ of these two methods under the same evolution time $t$. We define the fidelity $F = \langle \Psi_{\alpha,\pm}|\rho(t)|\Psi_{\alpha,\pm}\rangle$, where $\Psi_{\alpha,\pm} = N^{-1/2}(|\alpha\rangle \pm |-\alpha\rangle)$ are the target cat states that these two methods aim to prepare with $\alpha = \sqrt{4.3}$. The non-classical volume $\delta$ is another important factor to evaluate the prepared cat state, which is defined as \citep{kenfack2004negativity}
\begin{align}
\delta = \int |W(\alpha)|d^2\alpha - 1.
\end{align}
A higher non-classical volume $\delta$ indicates more apparent non-classical features \citep{wang2017hybrid}. Figure 7(a)
compares the fidelity $F$ of these two methods under the same  time $t$. At $t = 2.5/K$, the fidelity of TPC method is about $98\%$, which is much greater than the SPC method.  The comparison of the non-classical volumes $\delta$  of these two methods at the same evolution time is shown in Fig. 7(b). From $0$ to $t = 2.5/K$,  the non-classical volumes $\delta$ of TPC method increases rapidly, but $\delta$ of SPC method hardly changes. In the same evolutionary time $t$, the TPC method produces the cat state with a higher fidelity  and a larger  non-classical volume, which  means that it is faster. The fast preparation of cat sate by this TPC method is due to the fact that the minimum energy gap  $\Delta_{\text{min}}$ is 
increased by tuning  additional detuning $\Delta$. In addition, the minimum energy gap  $\Delta_{\text{min}}$ also affects the fidelity and  the non-classical volume of  prepared cat states. The larger  $\Delta_{\text{min}}$, the smaller  probability of the cat state transitioning to other excited states, and a higher  fidelity and a larger  non-classical volume cat state can be prepared.

\begin{figure}[!htbp]
	\centering
	\includegraphics[scale=0.35]{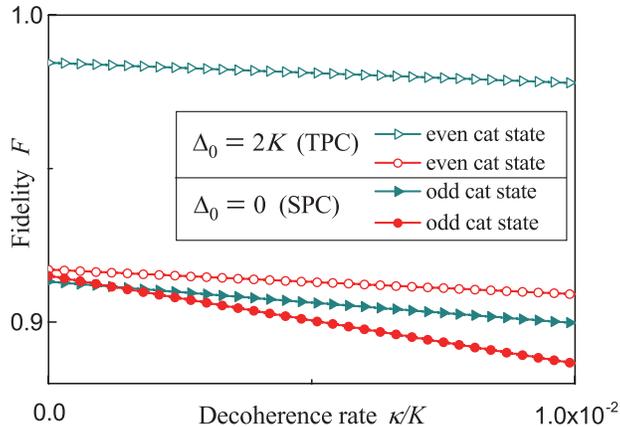}
    \caption{(Color online) The  fidelity $F$ of the cat states prepared by SPC method and TPC method as a function of the single-photon loss rate $\kappa/K$.  }	
\end{figure}

Up to now, the discussions is based on the unitary evolution without considering decoherence. In order to generalize to the open-system case, we introduce a Lindblad jump operator $L = \sqrt{\kappa}a$ to describe the single-photon loss, and the dynamics of the system is described by the master equation
\begin{align}\label{yangmei}
\dot{\rho} = -i[H, \rho] + L \rho L^{\dag} - \frac{1}{2}L^{\dag}L \rho - \frac{1}{2}\rho L^{\dag}L.
\end{align}
In the presence of single-photon loss, the cat state parity will  change randomly, reducing the fidelity of the prepared cat sate. We plot the  fidelity $F$ of the cat states prepared  by  TPC  and SPC methods  respectively as a function of single-photon loss $\kappa$.  The cat states prepared by these two methods are robust against single-photon loss very well. This is due to the fact that the two-photon driving $\beta$ can stabilize the cat state against the  rotation and dephasing  caused by Kerr nonlinearity \citep{puri2017engineering}. Here, we have not taken into account resonator dephasing and two-photon loss as they are typically negligible compared to single-photon loss \citep{puri2017engineering}.

\section{large scale cat states}\label{yff}
\begin{figure}[!htbp]
	\centering
	\includegraphics[scale=0.22]{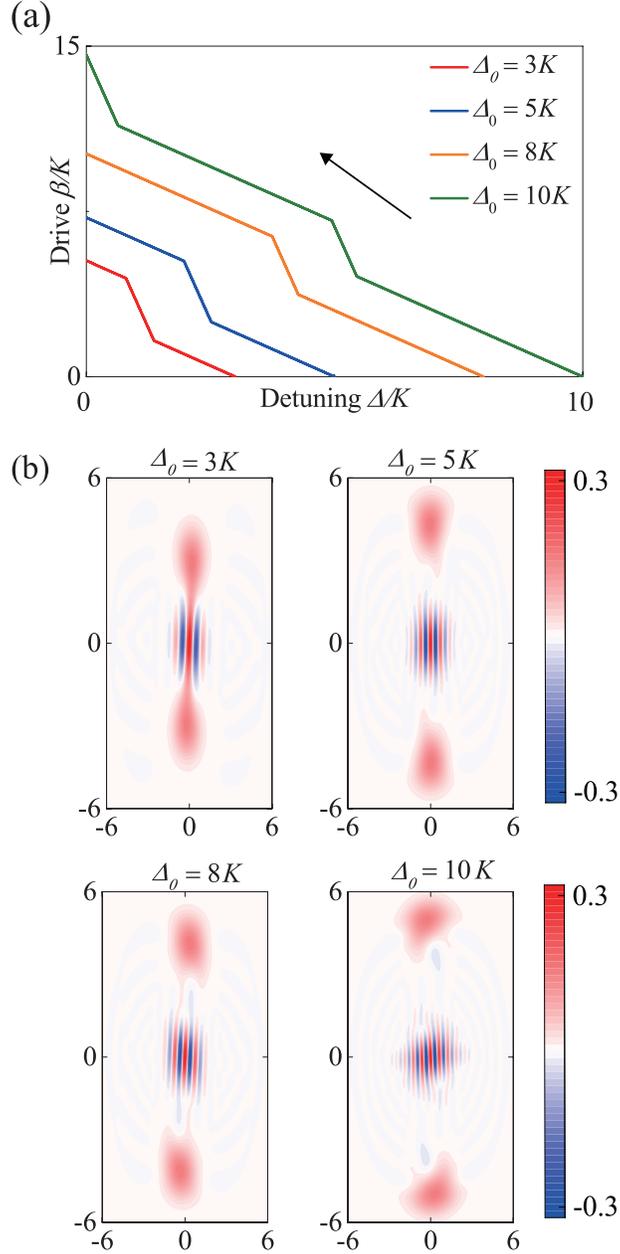}
    \caption{(a) Optimized path for generation cat states with different initial detuning $\Delta_0$. (b) Wigner functions $W(\alpha)$ of final state for KNR with different  initial detuning $\Delta_0$. }	
\end{figure}
Generally,  cat states with $|\alpha| \ge 2$ can be treated as good qubits. Because two  $|\pm \alpha\rangle$ of the cat state are not orthogonal  with each other and their overlap is determined by $|\langle \alpha|-\alpha\rangle| = \text{exp}(-2|\alpha|^2)$. For $|\alpha| \ge 2$, one has $|\langle \alpha|-\alpha\rangle| \approx 0$, and cat states  can be used as cat bits. Hence, the preparation of large-size cat states plays an important role in continuously variable quantum information processing.  Here, we will  investigate the effect on the size of the prepared cat states when the initial detuning $\Delta_0$ takes a larger value.  We adopt different   $\Delta_0$  and  discuss the effect of  $\Delta_0$  on the size of the prepared cat state. 

The optimal path for preparing cat states with different initial detuning $\Delta_0$ is shown in Fig. 9(a). We find that a large initial detuning $\Delta_0$ will lead to a large finial driving strength $\beta_f$. This means that the final Hamiltonian of the system can prepare  cat state with large size. We plot the  Wigner functions $W(\alpha)$ of the final cat states prepared by different initial $\Delta_0$ in Fig. 9(b)(the initial state is vacuum state $|0\rangle$).  The result shows that with the increase of  $\Delta_0$ , the distance $d$ between the two coherent states $|\pm \alpha \rangle$  gets farther, which also proves that the size of the cat state becomes larger.  The interference fringes of Wigner functions in Fig. 9(b) are also clear, which means that the  non-classical characteristics of  cat states are also well.
Therefore, we provide a method for preparing cat states with large-size.

\section{possible realization  with superconducting circuits}\label{doujiang}
\begin{figure}[!htbp]
	\centering
	\includegraphics[scale=0.26]{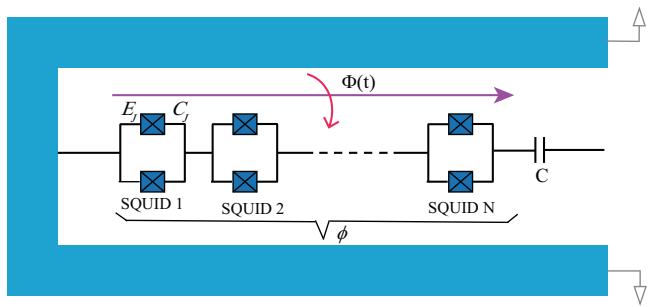}
    \caption{Schematic of a SQUID-array resonator with N SQUIDs. Each SQUID includes two identical Josephson junctions (with Josephson energy $E_J$ and  capacitance $C_J$). The SQUID array is shunted by  a shunting capacitor $C$, and the overall phase across the junction array is $\phi$. The external magnetic flux treading the SQUIDs is $\Phi(t)$.}	
\end{figure}
As shown in Fig. 10, we dicuss how to realize our proposal by considering  a SQUID-array resonator with N SQUIDs  which has implemented in
Refs. \citep{castellanos2007widely,castellanos2008amplification,wang2019quantum}. The  effective Hamiltonian of the system is given by \citep{wang2019quantum}
\begin{align}\label{heyao}
\hat{H} = 4E_C \hat{n}^2 - NE_J[\Phi(t)]\text{cos} \frac{\hat{\phi}}{N},
\end{align}
where $\hat{n}$ and $\hat{\phi}$ are the number of Cooper pairs and the overall phase across the junction array, respectively. Here, $E_C$ is the resonator's charging energy  including the contributions of the junction capacitance $C_J$ and the shunt capacitance $C$. $N$ is the number of SQUIDs in the array. $E_J$ is the Josephson energy for a single SQUID in the array, which depends on the external flux $\Phi(t)$.  The Josephson energy is periodically modulated by the external magnetic flux  $\Phi(t)$, which expressed as $E_J + \delta E_J \text{cos} \omega_p t$.

After Taylor expanding $\text{cos}(\hat{\phi}/N)$ in Eq. (\ref{heyao}) to fourth order, we obtain \citep{wang2019quantum,masuda2021controls}
\begin{align}
\begin{aligned}
\hat{H}/\hbar = \  &4E_C \hat{n}^2 - NE_J(1- \frac{1}{2}(\frac{\hat{\phi}}{N})^2 + \frac{1}{24}(\frac{\hat{\phi}}{N})^4 + \dots )\\&-N\delta E_J(1 - \frac{1}{2}(\frac{\hat{\phi}}{N})^2 + \dots) \text{cos} \omega_p t.
\end{aligned}
\end{align}
The quadratic time-independent part of the Hamiltonian can be diagonalized by defining
\begin{align}
\begin{aligned}
&\hat{n} = -in_0(\hat{a} - \hat{a}^{\dag}),\\&\hat{\phi} = \phi_0(\hat{a} + \hat{a}^{\dag}),
\end{aligned}
\end{align}
where $n^2_0 = \sqrt{E_J/32 N E_C}$ and $\phi^2_0 = \sqrt{2N E_C/E_J}$ are the zero-point fluctuations. After  quantization, we obtain the Hamiltonian of the SQUID array resonator
\begin{align}\label{qf}
\begin{aligned}
\frac{\hat{H}}{\hbar} = \ &\omega^{(0)}_c(\hat{a}^{\dag}\hat{a}+ \frac{1}{2}) + \frac{K}{6}(\hat{a}+\hat{a}^{\dag})^4 \\&+[-\frac{N\delta E_J}{\hbar} - 2\beta(\hat{a}+\hat{a}^{\dag})^2 - \frac{K\beta}{3\omega^{(0)}_c}(\hat{a}+\hat{a}^{\dag})^4]\text{cos} \omega_p t,
\end{aligned}
\end{align}
where $\omega_c^{(0)} = \sqrt{8E_C E_J/N}/\hbar$, $K = -E_C/N^2\hbar$ and $\beta = -\omega_c^{(0)}\delta E_J/8E_J$. Here,
$K$ and $\beta$ corresponds to the Kerr nonlinearity coefficient and the pump strength, respectively.   We consider a parameter regime in  Eq. (\ref{heyao}), where $\phi/N = 2\sqrt{2K/\omega_c^{(0)}}$ is sufficiently smaller than unity.   The last term in Eq. (\ref{qf}) can be neglected with an assumption, $K \beta \ll \omega_c^{(0)}$. Therefore, the  Hamiltonian of the SQUID array resonator is achieved \citep{masuda2021controls}
\begin{align}
\begin{aligned}
\hat{H}/\hbar =\ & \omega^{(0)}_c \hat{a}^{\dag}\hat{a} + \frac{K}{6}(\hat{a} + \hat{a}^{\dag})^4  -2\beta(\hat{a} + \hat{a}^{\dag})^2 \text{cos}\omega_p t.
\end{aligned}
\end{align}
By moving into a rotating frame at the frequency of $\omega_p/2$ and performing a rotating-wave approximation (neglect all the  rapidly oscillating terms),  an approximate Hamiltonian   can be obtained
\begin{align}
\hat{H}/\hbar = K \hat{a}^{\dag}\hat{a}^{\dag}\hat{a}\hat{a} + \Delta \hat{a}^{\dag}\hat{a} - \beta(\hat{a}^2 + \hat{a}^{\dag 2}),
\end{align}
where the detuning $\Delta = \omega^{(0)}_c - 2K - \omega_p/2$. This Hamiltonian has the same form as the Hamiltonian of Eq. (\ref{eq1}), which can be used to prepare cat states.

\section{conclusion}\label{yangsheng}
We  propose a quantum speedup method to  generate cat states in KNRs via  optimal adiabatic control sequences. This method is based on gradient-descent, searching in a two-parameter space $\{\Delta,\beta\}$, and finds an optimal evolution path through numerical calculation. This TPC method offers a practical way to speed up the  adiabatic evolution, which  offers significant improvements over the previous SPC method. The results show that the cat state prepared by TPC method has a higher fidelity $F$ and a larger non-classical volume $\delta$ than SPC method at the same evolution time $t$. The cat states prepared by TPC method are also robust against single-photon loss very well. In addition, this TPC method also  provides a promising solution to prepare a large-size cat state through a large initial detuning $\Delta_0$. We hope that these cat states can find wide applications in quantum information processing, especially in quantum information coding with cat states and quantum fault-tolerant computing.

\section*{Acknowledgments}
X.W. is supported by the China Postdoctoral Science Foundation No.2018M631136, and the National Science Foundation of China (Grant No. 11804270 and 12174303). H.R.L. is supported by the National Natural Science Foundation of China (Grant No.11774284).
\section*{References}

%\bibliography{bibliographycat}


%\end{thebibliography}%

\end{document}